\documentclass[twocolumn,showpacs,preprintnumbers,amsmath,amssymb,superscriptaddress]{revtex4}

\usepackage[dvipdfmx]{graphicx}
\usepackage{dcolumn}
\usepackage{bm}

\usepackage{subfigure}
\graphicspath{{fig/}}
\usepackage{tabularx}
\newcolumntype{Y}{>{\centering\arraybackslash}X} 
\newcolumntype{Z}{>{\raggedleft\arraybackslash}X} 
\DeclareFontFamily{U}{rsfs}{\skewchar\font127}
\DeclareFontShape{U}{rsfs}{m}{n}{<-6>rsfs5<6-8.5>rsfs7<8.5->rsfs10}{}
\DeclareSymbolFont{rsfs}{U}{rsfs}{m}{n}
\DeclareSymbolFontAlphabet{\mathrsfs}{rsfs}
\DeclareRobustCommand*\rsfs{\@fontswitch\relax\mathrsfs}

\begin{document}

\preprint{APS/123-QED}

\title{
Spin-charge-orbital ordering in hollandite-type manganites studied
by model Hartree-Fock calculation
}

\author{Makoto Fukuzawa}
\affiliation{Department of Physics, University of Tokyo, 5-1-5 Kashiwanoha, Kashiwa, 
Chiba 277-8561, Japan}
\author{Daiki Ootsuki}
\affiliation{Department of Physics, University of Tokyo, 5-1-5 Kashiwanoha, Kashiwa, 
Chiba 277-8561, Japan}
\author{Takashi~Mizokawa}
\affiliation{Department of Complexity Science and Engineering, 
University of Tokyo, 5-1-5 Kashiwanoha, Kashiwa, 
Chiba 277-8561, Japan}
\affiliation{Department of Physics, University of Tokyo, 5-1-5 Kashiwanoha, Kashiwa, 
Chiba 277-8561, Japan}

\date{\today}

\begin{abstract}
We investigate spin-charge-orbital ordering in a Mn$^{3+}$/Mn$^{4+}$ 
mixed valence state on a hollandite-type lattice using unrestricted 
Hartree-Fock calculation on a multi-band Mn 3$d$-O 2$p$ lattice model. 
The calculations show that the Mn$^{3+}$-Mn$^{4+}$ double exchange 
interaction, the Mn$^{3+}$-Mn$^{3+}$ and Mn$^{4+}$-Mn$^{4+}$ superexchange
interactions are ferromagnetic and play important roles to stabilize 
the charge and orbital ordering pattern. The most stable charge and orbital 
ordering pattern is consistent with the $1\times1\times1$ orthorhombic or 
monoclinic structure of K$_{1.6}$Mn$_8$O$_{16}$.
\end{abstract}

\maketitle

\newpage

\section{Introduction}

Various transition-metal oxides are known to show structural phase transitions 
which are accompanied by electronic transitions such as spin, charge, or orbital
orderings \cite{Mott1990,Tsuda2000}. Such phase transitions 
in perovskite-type transition-metal oxides including 
La$_{1-x}$Ca$_{x}$MnO$_3$ have been studied experimentally 
and theoretically, and the relationship between the structural 
transition and the charge-orbital ordering of transition-metal 
$d$ electrons has been revealed \cite{Imada1998}.
In the perovskite-type oxides, $M$O$_6$ ($M$ = transition metal)
octahedra share their corners and the electronic interaction
between the neighboring sites is dominated by the $M$-O-$M$ bond.
On the other hand, transition-metal oxides with edge-sharing
$M$O$_6$ octahedra including spinel-type Fe$_3$O$_4$ \cite{Fe3O4}
and rutile-type VO$_2$ \cite{VO2}, the direct $M$-$M$ bond also 
plays important role and often induces $M$-$M$ dimer formation.

Recently, novel structural transitions have been discovered 
in hollandite-type transition-metal oxides such as 
K$_2$V$_8$O$_{16}$ \cite{Isobe} and K$_2$Cr$_8$O$_{16}$. 
\cite{Hasegawa}
Since, in the hollandite-type structure, $M$O$_6$ octahedra share the corners 
and the edges, both the $M$-$M$ and $M$-O-$M$ bonds contribute to generate
spin-charge-orbital orderings. Therefore, the hollandite-type oxides 
have been inspiring efforts to develop new theoretical framework to 
describe possible mechanism of the structural transitions. 
For example, K$_2$V$_8$O$_{16}$ exhibits two step jumps of 
resistivity in the narrow temperature range around 170 K 
which correspond to two structural transitions from a high-temperature 
tetragonal structure to an intermediate-temperature tetragonal structure to 
a low-temperature monoclinic (almost orthorhombic) 
$\sqrt{2} \times \sqrt{2} \times 2$ structure. \cite{Isobe}
Since the formal valence of the V site is +3.75 for K$_2$V$_8$O$_{16}$,
it is expected that the metal-insulator transition and  
structural transitions are driven by charge ordering between V$^{3+}$ and V$^{4+}$
and the V$^{4+}$-V$^{4+}$ dimer formation along the V-O double chain
or the V-O ribon.
\cite{Horiuchi, Sakamaki, Mahadevan, Komarek, Brink}
On the other hand, the structural transition in K$_2$Cr$_8$O$_{16}$ 
is well described as a Peierls transition of the itinerant
Cr 3$d$ $t_{2g}$ electrons. \cite{Toriyama}

Compared to K$_2$V$_8$O$_{16}$ and K$_2$Cr$_8$O$_{16}$, 
the physical properties of K$_2$Mn$_8$O$_{16}$ are not well 
understood yet. First of all, the concentration of K ions
tends to be reduced, and the actual composition is close to
K$_{1.6}$Mn$_8$O$_{16}$. \cite{Kuwabara}
At 380 K, K$_{1.6}$Mn$_8$O$_{16}$ undergoes a structural phase 
transition from a high-temperature tetragonal phase 
to a low-temperature monoclinic (almost orthorhombic) 
$1\times1\times 1$ phase. 
At 250 K, K$_{1.6}$Mn$_8$O$_{16}$ exhibits another
phase transition to a monoclinic $1\times5\times 1$ 
superstructure phase with five-fold periodicity along 
the Mn-O double chain or the Mn-O ribon. \cite{Kuwabara}
In case of K$_{x}$Mn$_8$O$_{16}$, the filling of K$^{+}$ ions
is (2-$x$)/2, and the ratio between Mn$^{3+}$ and Mn$^{4+}$
is $x$/8 : (8-$x$)/8. Therefore, assuming that K$^{+}$ is located
at neighbors of Mn$^{3+}$, the K$^{+}$ ion order and the Mn$^{3+}$/Mn$^{4+}$
charge order can collaborate only when (2-$x$):$x$ = $x$:(8-$x$), namely $x$=1.6.

In the present study, we focus on the origin of the phase transition at 380 K
from the tetragonal phase to the low-temperature monoclinic phase without 
the superstructure. Since the unit cell of K$_x$Mn$_8$O$_{16}$ contains 
eight Mn sites, Mn 3$d$ charge and orbital ordering can occur without
superstructure. In particular, since the ratio between Mn$^{3+}$ and Mn$^{4+}$
is 1/4 : 3/4 for $x$=2, the charge ordering at $x$=2 is expected to be compatible 
with the unit cell.
We have examined possible charge and orbital orderings using 
unrestricted Hartree-Fock calculation on a hollandite-type multi-band 
Mn 3$d$-O 2$p$ lattice model. 
The calculations show that the Mn$^{3+}$-Mn$^{4+}$ double exchange 
interaction and the Mn$^{3+}$-Mn$^{3+}$ and Mn$^{4+}$-Mn$^{4+}$ 
superexchange interaction are ferromagnetic and play important roles 
to stabilize the charge and orbital ordering pattern which is consistent 
with the $1\times1\times 1$ orthorhombic or monoclinic structure.

\section{ Method}
We carried out unrestricted Hartree-Fock calculation for two layers of neighboring 
two double-chain tunnels where 32 Mn sites and 64 O sites are considered. 
We employ the multiband $d$-$p$ model where full degeneracy of the Mn $3d$ orbitals 
and O $2p$ orbitals are taken into account. \cite{Mizokawa96}
The Hamiltonian is given by
\begin{align}
\hat{\mathrsfs{H}} =& \hat{\mathrsfs{H}}_p + \hat{\mathrsfs{H}}_d + \hat{\mathrsfs{H}}_{pd}  \notag \\
 \hat{\mathrsfs{H}}_p =& \sum_{kl\sigma} \epsilon^p_k p^{\dagger}_{kl\sigma} p_{kl\sigma} + \sum_{kll'\sigma}V^{pp}_{kll'}p^{\dagger}_{kl\sigma} p_{kl'\sigma} + \text{h.c.} \notag \\
\hat{\mathrsfs{H}}_d =& \epsilon^0_d \sum_{i \alpha m \sigma} d^{\dagger}_{i \alpha m \sigma}d_{i \alpha m \sigma} + \sum_{i \alpha mm'\sigma \sigma'}h_{mm'\sigma \sigma'}d^{\dagger}_{i \alpha m \sigma}d_{i \alpha m' \sigma'}    \notag \\
&+ u\sum_{i \alpha m} d^{\dagger}_{i \alpha m \uparrow}d_{i \alpha m \uparrow} d^{\dagger}_{i \alpha m \downarrow}d_{i \alpha m \downarrow}  \notag \\
& +u'\sum_{i \alpha m m'} d^{\dagger}_{i \alpha m \uparrow}d_{i \alpha m \uparrow}  d^{\dagger}_{i \alpha m \downarrow}d_{i \alpha m \downarrow} \notag \\
&+ (u'-j)\sum_{i \alpha mm'\sigma}  d^{\dagger}_{i \alpha m \sigma}d_{i \alpha m \sigma}  d^{\dagger}_{i \alpha m' \sigma}d_{i \alpha m' \sigma}  \notag \\
&+ j \sum_{i \alpha mm'} d^{\dagger}_{i \alpha m \uparrow} d_{i \alpha m' \uparrow} d^{\dagger}_{i \alpha m' \downarrow}d_{i \alpha m \downarrow}  \notag \\
&+ j' \sum_{i \alpha mm'} d^{\dagger}_{i \alpha m \uparrow} d_{i \alpha m' \uparrow}  d^{\dagger}_{i \alpha m \downarrow}d_{i \alpha m' \downarrow} \notag \\
\hat{\mathrsfs{H}}_{pd} =& \sum_{kml\sigma}V^{pd}_{kml}d^{\dagger}_{km\sigma} p_{kl\sigma} + \text{h.c.} \notag 
\end{align}

Here, $d^{\dagger}_{i \alpha m \sigma}$ are creation operators for the Mn $3d$ electrons at site $\alpha$ 
of the $i^{\text{th}}$ unit cell and $d^{\dagger}_{km\sigma}$ and $p^{\dagger}_{kl\sigma}$ are creation 
operators for Bloch electrons which are constructed from the $m^{\text{th}}$ component of the Mn $3d$ 
orbitals and from the $l^{\text{th}}$ component of the O $2p$ orbitals, respectively, with wave vector $\bm{k}$. 
The matrix $h_{mm'\sigma \sigma'}$ represents the crystal field splitting. 
The transfer integrals between the O $2p$ orbitals $V^{pp}_{kll'}$ are given by Slater-Koster parameters 
($pp\sigma$) and ($pp\pi$) which are fixed at $0.60$ eV and $-0.15$ eV respectively. 
The transfer integrals between the Mn $3d$ and O $2p$ orbitals $V^{pd}_{kml}$ are represented 
by ($pd\pi$) and ($pd\sigma$). They are fixed as ($pd\sigma$) = -2.0 eV and ($pd\pi$)= 0.9 eV.
Kanamori paramters $u$, $u'$, $j$, and $j'$ satisfies $u=u'+j+j'$ and $j'=j$.
$u$ and $j$ are fixed at 7.3 eV and 0.8 eV, respectively.
The O 2$p$-to-Mn 3$d$ charge transfer energy is $\Delta=\epsilon^d-\epsilon^p+nU$ 
where $U=u-20j/9$ and $n$ is the number of Mn 3$d$ electrons. 
$\Delta$ is set to 1.44 eV in the present calculation which is close to 
typical $\Delta$ values for Mn$^{4+}$ oxides. \cite{Wadati}

\section{\label{sec:RaD}Results and Discussion}

\begin{figure}[htbp]
  \includegraphics[clip,width=10cm,height=10cm,keepaspectratio]{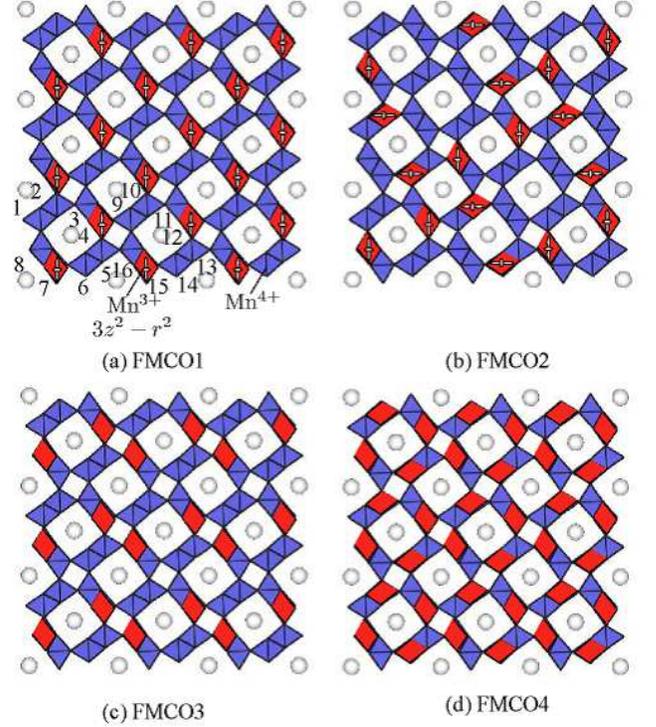}
  \caption{(color online)
    Charge ordering patterns for ferromagnetic states with 
    Mn$^{3+}$ : Mn$^{4+}$ = 1/4 : 3/4
    which are labelled as (a) FMCO1, (b) FMCO2, (c) FMCO3(c), and (d) FMCO4.
  }
  \label{fig:states}
\end{figure}

The unrestricted Hartree-Fock analysis for K$_2$Mn$_8$O$_{16}$ 
with the reasonable parameter set provides ferromagnetic solutions 
with several charge ordering patterns. The charge ordering patterns
are illustrated in Fig. \ref{fig:states}.
The energies of those states are listed in Table \ref{tab:1}.

\begin{table}[htbp]
  \begin{center}
    \caption{Energy per unit cell of the FMCO2, FMCO3, and FMCO4 states
      relative to the most stable FMCO1 state.}
      \begin{tabularx}{\columnwidth}{YYYYYYY} \hline\hline
        State & FMCO2 & FMCO3 & FMCO4  \\ \hline
        Energy (eV) & 0.008 & 0.252 & 0.276  \\ \hline\hline
      \end{tabularx}
    \end{center}
  \label{tab:1}
\end{table}

\begin{figure}[htbp]
  \centering
  \includegraphics[clip,width=\linewidth,keepaspectratio]{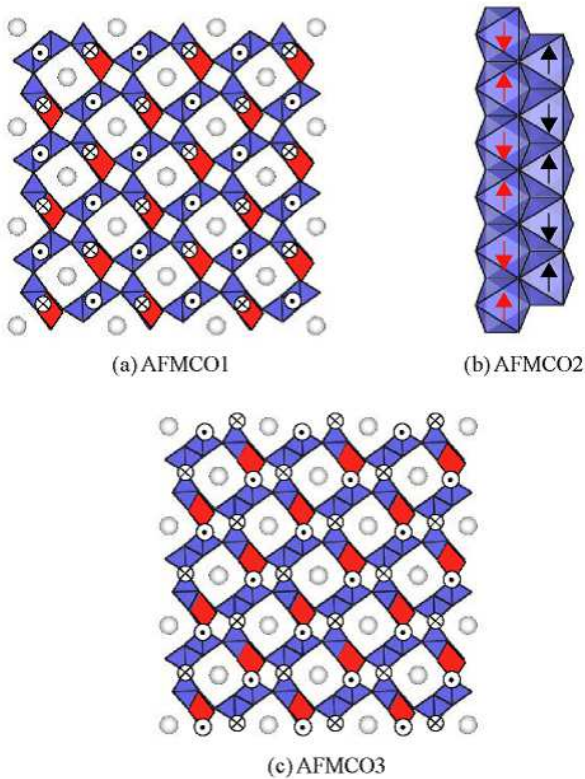}
  \caption{
    (color online)Charge ordering patterns for antiferromagneic states
    with  Mn$^{3+}$ : Mn$^{4+}$ = 1/4 : 3/4
    which are labeled as (a) AFMCO1, (b) AFMCO2, and (c) AFMCO3.
    The closed circles and crosses indicate the spin up and
    down sites, respectively. 
  }
  \label{fig:Antiferro}
\end{figure}

\begin{table}[htbp]
  \begin{center}
    \caption{Energy per unit cell of the AFMCO2, AFMCO3, and AFMCO4 
      states relative to the most stable FMCO1 state.}
    \begin{tabularx}{\columnwidth}{YYYYYYY} \hline\hline
      State & AFMCO1 & AFMCO2 & AFMCO3 \\ \hline
      Energy (eV) & 3.045 & 1.448 & 0.965 \\ \hline\hline
    \end{tabularx}
  \end{center}
  \label{tab:2}
\end{table}

\begin{figure}[htbp]
  \centering
  \includegraphics[clip,width=\linewidth,keepaspectratio]{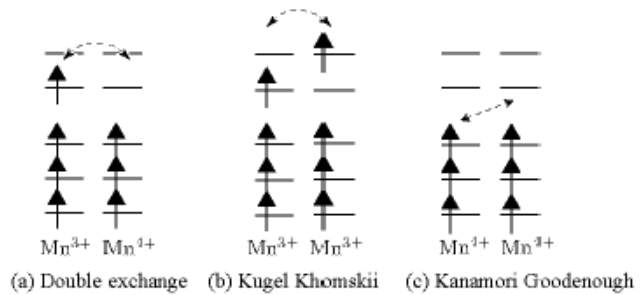}
  \caption{Schematic pictures for (a) Mn$^{3+}$-Mn$^{4+}$
double exchange interaction,
(b) Mn$^{3+}$-Mn$^{3}$ Kugel-Khomskii superexchange interaction,
and(c) Mn$^{4+}$-Mn$^{4}$ Goodenough-Kanamori 
superexchange interaction.}
  \label{fig:exchange}
\end{figure}	

The most stable state is the FMCO1 state which is shown in Fig.~\ref{fig:states}(a). 
Here, the charge ordering pattern in the a-b plane is illustrated.
There are two kinds of double chains running along the c axis in the FMCO1 state.
The one consists of only Mn$^{4+}$O$_{6}$ octahedra, and the other consists of 
Mn$^{3+}$O$_{6}$ and Mn$^{4+}$O$_{6}$ octahedra. 
In the latter type of double chain, the Mn$^{3+}$ and Mn$^{4+}$ sites are aligned 
along the c axis in straight lines, respectively. 
As for the orbital ordering, at the Mn$^{3+}$ sites with one $e_g$ electron, 
the Mn 3$d$ $3z^{2}-r^{2}$ orbital [indicated by the cigar-like orbital shape
in Fig.~\ref{fig:states}(a)] is directed to the corner oxygen which is
sandwiched by the Mn$^{3+}$ and Mn$^{4+}$ sites. This situation is very similar
to the Mn$^{3+}$/Mn$^{4+}$ mixed valence perovskite-type Mn oxides. \cite{Mizokawa97}

Figures~\ref{fig:states}(b)-(d) show other three kinds of ferromagnetic solutions 
with reasonable symmetries. The FMCO2 state is similar to the FMCO1 state. 
The Mn$^{3+}$ and Mn$^{4+}$ sites are aligned along the c axis in straight lines.
However, the charge ordering pattern in the a-b plane keeps the tetragonal symmetry.
In the FMCO3 state, the charge ordering pattern in the a-b plane is
the similar to (but somewhat different from ) the FMCO1 and FMCO2 states
but the Mn$^{3+}$ and Mn$^{4+}$ ions stack alternately 
along the double chains or the c axis. 
In the FMCO4 state, one plane has the charge ordering pattern with 
Mn$^{3+}$ : Mn$^{4+}$ = 1/2 : 1/2 and the next plane is filled with Mn$^{4+}$.
These states are higher in energy than the FMCO1 state (see Table.~\ref{tab:1}).

Since K$_{1.6}$Mn$_8$O$_{16}$ becomes antiferromagetic 
at low temperature experimentally, we also investigated 
antiferromagnetic solutions of the present model 
calculation to elucidate the magnetic property of the system.
Figure~\ref{fig:Antiferro} shows three different types of
antiferromagnetic configurations. In the AFMCO1 state, 
the Mn $3d$ spins are antiferromagnetic between 
corner-sharing octahedra as shown in Fig.~\ref{fig:Antiferro}(a).
The charge-ordering pattern of the AFMCO1 state is the same as 
the most stable FMCO1 state, and both the FMCO1 and AFMCO state
have the ferromagnetic double chains with Mn$^{4+}$O$_{6}$ octahedra and 
with Mn$^{3+}$O$_{6}$ and Mn$^{4+}$O$_{6}$ octahedra.
Therefore, the energy difference between the FMCO1 and AFMCO1 states
is due to the Mn$^{3+}$- Mn$^{4+}$ double exchange interaction 
between the corner-sharing MnO$_6$ octahedra or between the neighboring double chains.
In the AFMCO2 state, the Mn $3d$ spins are antiferromagnetic 
between the edge-sharing octahedra in the double chain along 
the c axis [see Fig.~\ref{fig:Antiferro}(b)], whereas the spin and
charge arrangement in the a-b plane of the AFMCO2 is the same as 
the most stable FMCO1 state. 
In the AFMCO3 state, the Mn $3d$ spins are antiferromagnetic 
between the edge-sharing octahedra in the double chain along
the a-b plane [see Fig.~\ref{fig:Antiferro}(c)], whereas the
ferromagnetic coupling between the corner-sharing MnO$_6$ octahedra
and that along the c-axis are the same as the most stable FMCO1 state.
Compared to the ferromagnetic states, all the obtained antiferromagnetic states 
have much higher energy values (see Table.~\ref{tab:2}). 
In the present calculation, antiferromagnetic coupling between 
far-distant-neighbors is not considered. We speculate that 
the antiferromagetic state of K$_{1.6}$Mn$_8$O$_{16}$ is 
of helical type stabilized by nearest neighbor 
ferromagnetic coupling and far-distant-neighbor 
antiferromagnetic coupling.
The AFMCO2 and AFMCO3 states are much lower in energy than 
the AFMCO2 state, indicating that the ferromagnetic double exchange
interaction between the corner-sharing MnO$_6$ octahedra 
is much stronger than the ferromagnetic coupling between 
the edge-sharing octahedra.

\begin{figure}[htbp]
  \centering
  \subfigure[488 K$_{2}$Mn$_{8}$O$_{16}$]{%
    \includegraphics[clip,width=4.2cm]{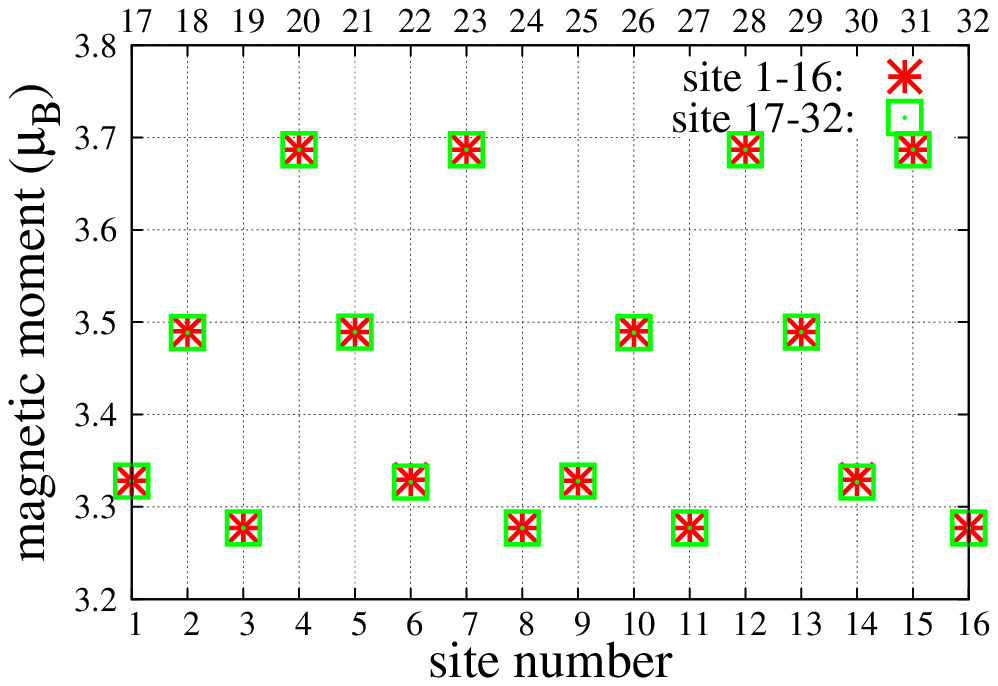}}
  \subfigure[487 K$_{1.75}$Mn$_{8}$O$_{16}$]{%
    \includegraphics[clip,width=4.2cm]{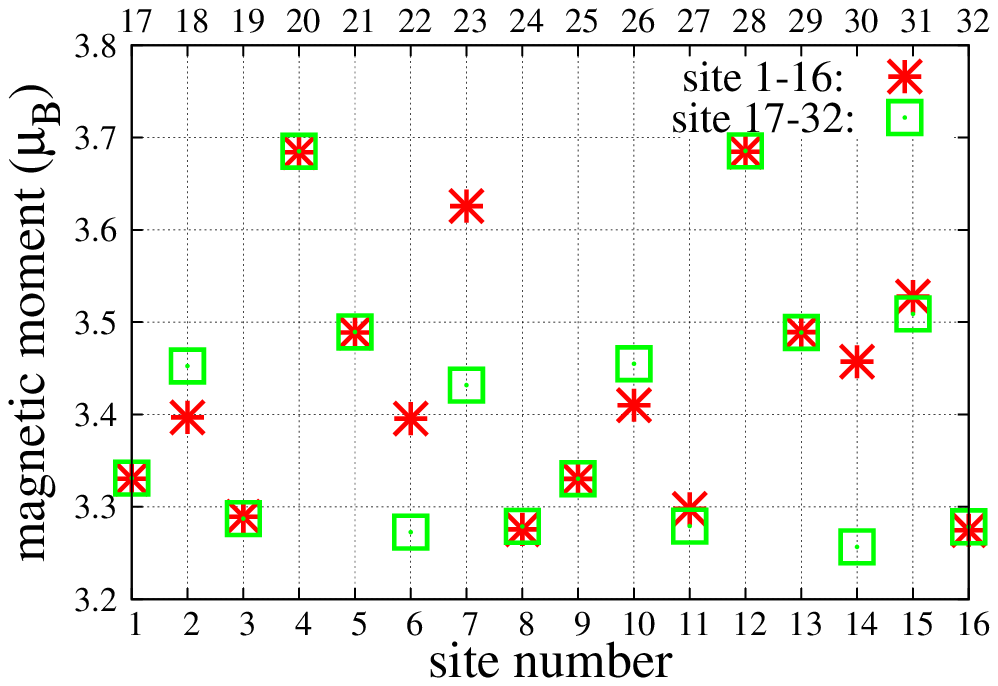}}
  \subfigure[486 K$_{1.5}$Mn$_{8}$O$_{16}$(not converged)]{%
    \includegraphics[clip,width=4.2cm]{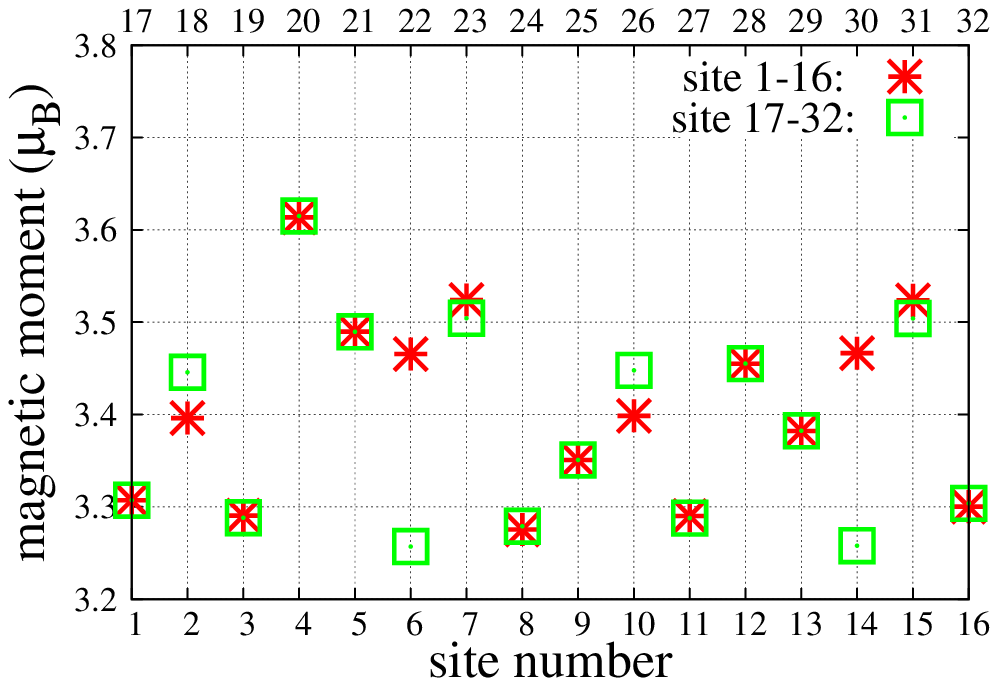}}
  \subfigure[485 K$_{1.25}$Mn$_{8}$O$_{16}$]{%
    \includegraphics[clip,width=4.2cm]{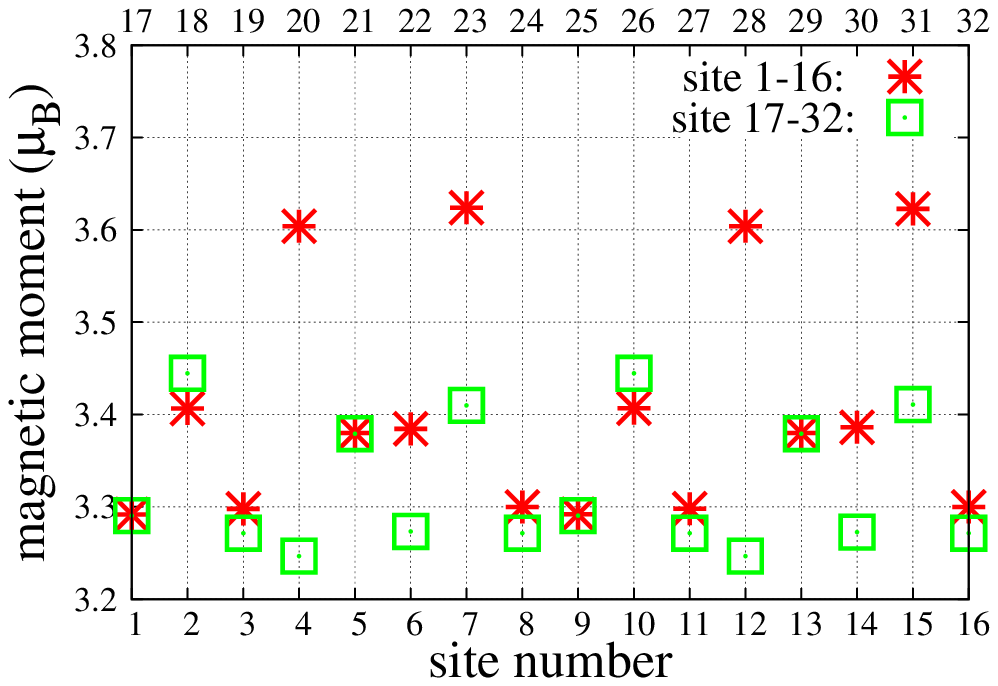}}
  \subfigure[484 KMn$_{8}$O$_{16}$]{%
    \includegraphics[clip,width=4.2cm]{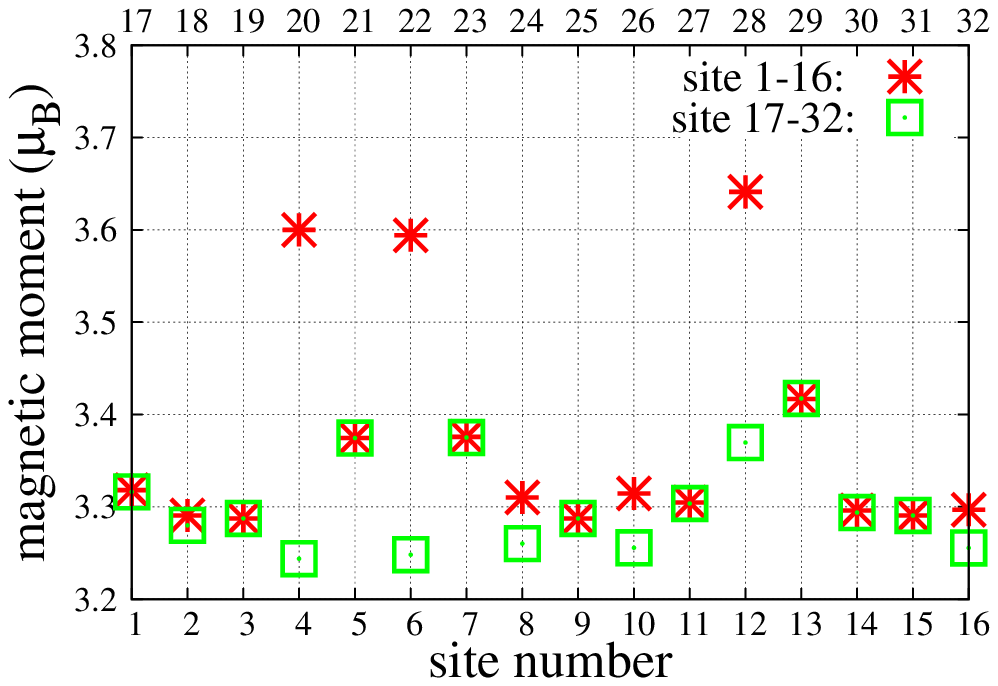}}
  \subfigure[483 K$_{0.75}$Mn$_{8}$O$_{16}$]{%
    \includegraphics[clip,width=4.2cm]{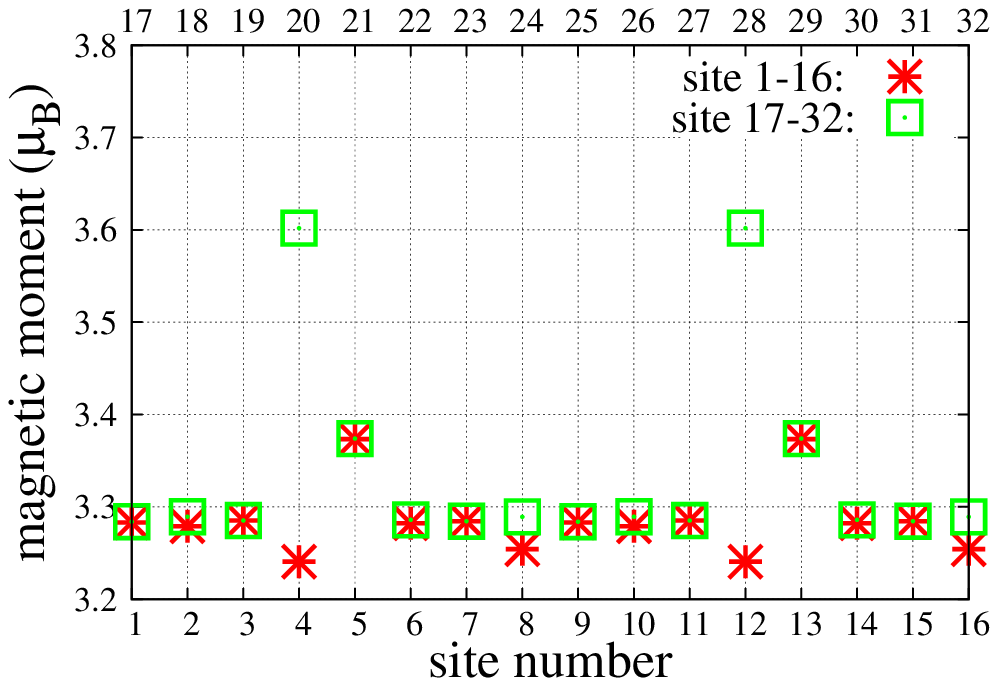}}
  \subfigure[482 K$_{0.5}$Mn$_{8}$O$_{16}$]{%
    \includegraphics[clip,width=4.2cm]{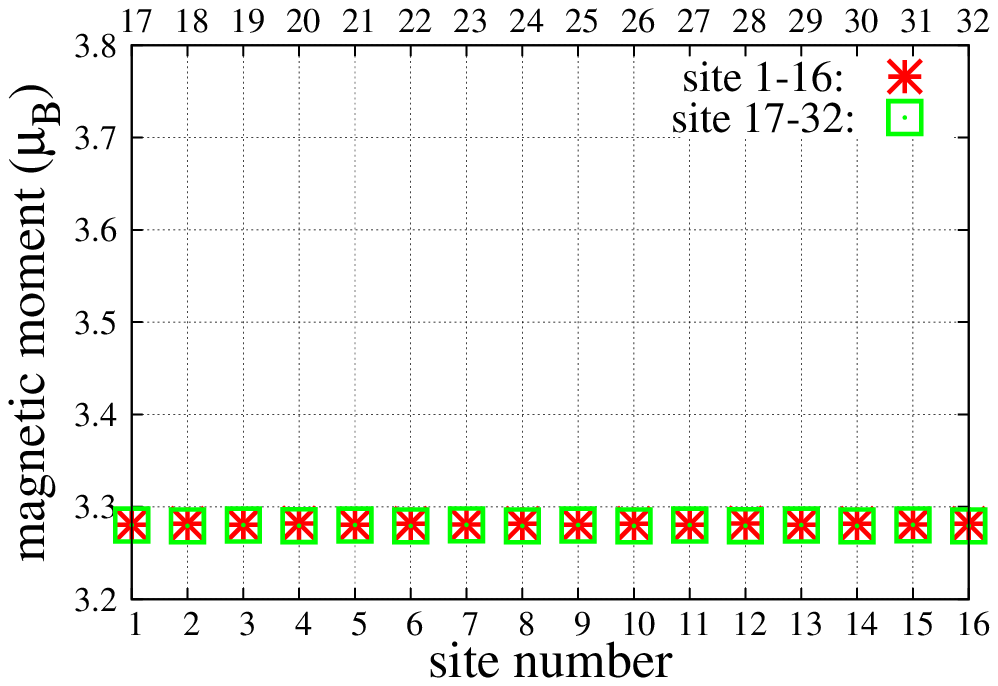}}
  \subfigure[481 K$_{0.25}$Mn$_{8}$O$_{16}$]{%
    \includegraphics[clip,width=4.2cm]{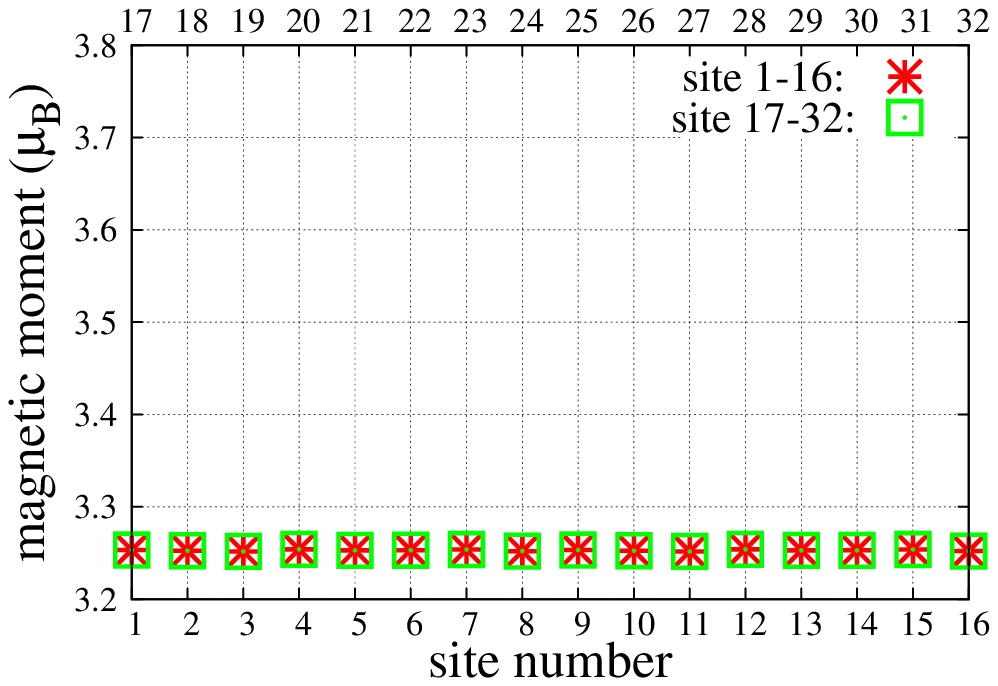}}
  \caption{(color online)Magnitudes of Mn 3$d$ spins for 32 Mn sites 
in the FMCO1 state obtained in the present model 
calculation. Hole doping level $x$ is set to (a) $x$=0.0,
(b) $x$=0.25, (c) $x$=0.5, (d) $x$=0.75, (e) $x$=1.0, 
(f) $x$=1.25, (g) $x$=1.5, and (h) $x$=1.75.
}
  \label{fig:holedopeing}
\end{figure}

At this stage, we discuss the origin of the spin-charge-orbital ordering
of the FMCO1 state. As a reason of ferromagnetic coupling between the Mn spins, 
three types of electronic exchange interactions are possible for this system
which are shown schematically in Fig.~\ref{fig:exchange}. 
The Mn$^{3+}$-Mn$^{4+}$ ferromagnetic coupling is derived from the double exchange interaction 
which can be enhanced by the orbital ordering of the Mn$^{3+}$ site.
The Mn$^{3+}$-Mn$^{3+}$ ferromagnetic coupling (along the c axis) is induced by the superexchange 
interaction with orbital degeneracy, which is explained by Kugel-Khomskii mechanism. 
As for the Mn$^{4+}$-Mn$^{4+}$ superexchange interaction with the Mn-O-Mn bond in the $90^{\circ}$ 
angle configuration, the ferromagnetic coupling is explained by Kanamori-Goodenough rule.
Among the antiferromagnetic states, the AFMCO3 state is more stable
than the AFMCO1 and AFMCO2 states, which indicates  that the double
exchange type ferromagnetic interaction between corner sharing
MnO$_{6}$ octahedra is the strongest and the Mn$^{3+}$-Mn$^{3+}$ 
ferromagnetic coupling along the c axis is the second strongest.
The difference in the Mn$^{3+}$-Mn$^{4+}$ double exchange interaction
should be responsible for the small energy difference between the FMCO1
and FMCO2 states.
The number of Mn 3$d$ spin at each Mn site of the FMCO1 state 
is shown in Fig.~\ref{fig:holedopeing}(a).
The 16 Mn sites in the first layer are labelled as 1-16 which is shown in Fig.~\ref{fig:states}(a).
The remaining 16 Mn sites in the second layer are labelled as 17-32. 
The Mn$^{3+}$ sites have $\sim$ 3.7 $\mu_B$.
Whereas most of the Mn$^{4+}$ sites have $\sim$ 3.3 $\mu_B$,  
the Mn$^{4+}$ sites sharing the corner oxygens 
with the Mn$^{3+}$ sites have $\sim$ 3.5 $\mu_B$. 
The increase of the Mn 3$d$ spins in the Mn$^{4+}$ sites
is due to the leakage of the Mn 3$d$ $e_g$ spins 
by the double exchange coupling.
Since the distance between the neighboring Mn$^{3+}$ chains is shorter in the FMCO2 state
than that in the FMCO1 state, the energy gain by the double exchange interaction can be slightly 
larger in the FMCO1 state. 
In the FMCO1 state, the orbital ordering of the Mn$^{3+}$ site
contributes to enhance the double exchange interaction 
between the Mn$^{3+}$ and Mn$^{4+}$ sites.
This charge-orbital ordering pattern of the FMCO1 state does not have 
tetragonal symmetry whereas it is compatible with the unit cell of K$_2$Mn$_8$O$_{16}$.
Therefore, the FMCO1 state with $1\times1\times1$ unit cell is consistent 
with the intermediate phase of K$_{1.6}$Mn$_8$O$_{16}$ realized between 250 K and 380 K.

\begin{table}[htbp]
  \centering
  \caption{Hole doping dependence of band gap of the FMCO1 state
    for K$_{2-x}$Mn$_{8}$O$_{16}$.}
  \begin{tabular}{lcccccccc} \hline\hline
    $x$ & 0.0 & 0.25 & 0.5 & 0.75 & 1.0 & 1.25 & 1.5 & 1.75 \\ \hline
    Gap (eV) & 0.419 & 0.130 & 0.039 & 0.120 & 0.060 & 0.120 & 0.00 & 0.00 \\ \hline\hline
  \end{tabular}
  \label{tab:3}
\end{table}

In the next step, we examine the stability of the FMCO1 state against 
the reduction of K content or the hole doping to the Mn$_8$O$_{16}$ lattice.
Figure~\ref{fig:holedopeing} and Table~\ref{tab:3} show the hole doping effect 
on the system where the reference point is set to the FMCO1 state of K$_2$Mn$_8$O$_{16}$. 
The total number of the Mn 3$d$ and O 2$p$ electrons for K$_{2}$Mn$_{8}$O$_{16}$ is 488
in the present model, and the hole doping corresponds to reduction of K atoms from the system. 
The total number of the Mn 3$d$ and O 2$p$ electrons is $488-x$ in the present model
for K$_{2-0.25x}$Mn$_{8}$O$_{16}$. 
As shown in Fig.~\ref{fig:holedopeing}, the charge ordering pattern remains in going from
$x$=2 to $x$=1, indicating that the FMCO1 state obtained for K$_2$Mn$_8$O$_{16}$ is 
relevant for K$_{1.6}$Mn$_8$O$_{16}$. The charge ordering pattern is slightly disturbed 
by the hole doping which can couple with the superstructure along the c-axis due to
the K vacancy ordering and would be an origin of far-distant-neighbor antiferromagnetic coupling.
As shown in Table~\ref{tab:3}, the magnitude of the band gap tends to decrease 
as the amount of hole increases until the transition to a metallic state 
at K$_{0.5}$Mn$_{8}$O$_{16}$. 
It should be noted that in the states K$_{1.75}$Mn$_{8}$O$_{16}$,
K$_{1.25}$Mn$_{8}$O$_{16}$, and KMn$_{8}$O$_{16}$, the Mn 3$d$ $e_g$ level is
partially occupied at the "Mn$^{4+}$" sites sharing the corner oxygens 
with the Mn$^{3+}$ sites and, these states have relatively wide gaps, 
which we infer results from the double exchange interaction between 
Mn$^{3+}$ and Mn$^{4+}$.

\section{Conclusion}
In conclusion, we investigate spin-charge-orbital ordering in a Mn$^{3+}$/Mn$^{4+}$ 
mixed valence state on a hollandite-type lattice using unrestricted 
Hartree-Fock calculation on a multi-band Mn 3$d$-O 2$p$ lattice model. 
The Mn$^{3+}$-Mn$^{4+}$ ferromagnetic coupling due to the double exchange interaction 
plays essential role to stabilize the charge ordering pattern.
In addition, The Mn$^{3+}$-Mn$^{3+}$ and Mn$^{4+}$-Mn$^{4+}$ ferromagnetic couplings
along the c axis is induced by the Kugel-Khomskii and  Kanamori-Goodenough mechanisms. 
The most stable charge and orbital ordering pattern is consistent with the 
$1\times1\times1$ orthorhombic or monoclinic structure realized in K$_{1.6}$Mn$_{8}$O$_{16}$.

\section*{Acknowledgement}
The authors would like to thank Prof. Y. Ueda, Dr. M. Isobe, and Prof. D. I. Khomskii 
for valuable discussions. 
M. F. is supported by Japan Society for the Promotion of Science through 
Program for Leading Graduate Schools (MERIT).

\end{document}